\documentclass[preprint]{revtex4}
\usepackage{graphicx}
\usepackage{amsmath}
\usepackage{amsfonts}

\begin{document}

\title{Guided atom laser : a new tool for guided atom optics} 

\author{J. Billy}
\email{juliette.billy@institutoptique.fr}
\homepage{http://www.atomoptic.fr}
\affiliation{Laboratoire Charles Fabry de l'Institut d'Optique, 
CNRS and Univ. Paris XI, 
Campus Polytechnique, 2 av A. Fresnel, 91128 PALAISEAU cedex, France}
\author{V. Josse}
\affiliation{Laboratoire Charles Fabry de l'Institut d'Optique, 
CNRS and Univ. Paris XI, 
Campus Polytechnique, 2 av A. Fresnel, 91128 PALAISEAU cedex, France}
\author{Z. Zuo}

\author{W. Gu\'erin}
\altaffiliation{Institut Non LinŽaire de Nice (UMR 6618 CNRS)
1361, route des Lucioles, 06560 Valbonne, France}
\affiliation{Laboratoire Charles Fabry de l'Institut d'Optique, 
CNRS and Univ. Paris XI, 
Campus Polytechnique, 2 av A. Fresnel, 91128 PALAISEAU cedex, France}
\author{A. Aspect}
\affiliation{Laboratoire Charles Fabry de l'Institut d'Optique, 
CNRS and Univ. Paris XI, 
Campus Polytechnique, 2 av A. Fresnel, 91128 PALAISEAU cedex, France}
\author{P. Bouyer}
\affiliation{Laboratoire Charles Fabry de l'Institut d'Optique, 
CNRS and Univ. Paris XI, 
Campus Polytechnique, 2 av A. Fresnel, 91128 PALAISEAU cedex, France}

\date{\today}
\begin{abstract} 

We present a guided atom laser. A Bose-Einstein condensate (BEC) is created in a crossed
hybrid magnetic and an elongated optical trap, which acts as a matterwave guide. Atoms are extracted from the BEC
by radio frequency (rf) outcoupling and then guided in the horizontal optical matterwave guide. This method allows to
control the acceleration of the beam and to achieve large de~Broglie wavelength. We also measure the longitudinal
energy of the guided atom laser using atom optical elements based on a blue light barrier.

\end{abstract}
\maketitle
%
\section{Introduction}
The field of Bose-Einstein condensation of atomic
vapors is undergoing a rapid experimental development, providing
a rich new phenomenology. Along these lines, the possibility of building atom
interferometers \cite{lecoq:2005,Gupta:2002} and guiding condensed particles \cite{Leanhardt:2002,Bongs:2000}
through various geometries
opens up the prospects of a rich variety of interference \cite{Wang:2005,Shin:2005},
transport \cite{Hansel:2001/2,Clement:2005}, and/or coherence phenomena \cite{Hansel:2001/3}. 
The occupation of a single quantum 
state by a large number of identical bosons is a 
matter-wave analog to the storage of photons in a single mode of a laser cavity. Just as one extracts
a coherent, directed beam of photons from a laser cavity by using a partially transmitting
mirror as an output coupler, one can analogously extract directed matter waves
from a condensate \cite{Ketterle:1997,Hagley:1999,Bloch:1999}
which are now used as a coherent sources atoms.  
Such atom laser is important in the
field of atom optics, the manipulation of atoms being analogous to the manipulation of light.
Its development is providing atom sources that are as different from ordinary atomic
beams as lasers are from light bulbs. 
 In parallel to the development of atom lasers, there are numerous efforts for miniaturizing the structures
used to confine the atoms \cite{Weinstein:1995,Thywissen:1999}. Following the successful trapping
and guiding of thermal atoms using self-supported
miniature wires \cite{Fortagh:1998,Denschlag:1999,Key:2000,Teo:2001,Cren:2002} and substrate-supported microfabricated
wire arrays \cite{Reichel:1999,Muller:1999,Dekker:2000,Folmann:2000}, recent experiments merged wire
traps on the millimeter scale and microfabricated electronic
devices \cite{Ott:2001,Hansel:2001} with Bose-Einstein condensation.

As for classical optical sources, the transfer of cold atoms from magneto-optic traps
into this small atom guides represents a problem
because the source volume of the order 1 mm$^3$ cannot be accommodated inside small atom guides. Hence,
the coupling of cold atoms into a microscopic atomic waveguide has been accomplished by injecting fast moving atoms
(10 m/s) from a low velocity source such as LVIS or 2D-MOT
\cite{Lu:1996,Muller:1999}. While this approach allows for a high flux of atoms
coupled into the atom guide, the large temperatures of the
injected atoms pose a disadvantage for future atom interferometry
experiments. Alternatively, slower atoms can
be coupled into an atom guide by placing the entrance of the
guide vertically below a MOT and dropping the atoms into
the guide using gravity \cite{Dekker:2000,Key:2000}. In this approach, the expansion
of the atom cloud during the free fall leads to a low
coupling efficiency. Although many
cold atoms sources have already
been injected in various guides \cite{Denschlag:1999,Dekker:2000,Key:2000,Muller:2000}
better efficiency can be achieved with coherent sources
of atoms which have recently been designed using various output
coupling schemes. Nevertheless,  even though continuous guided
beams of condensed atoms will be accessible in the near
future  \cite{Mandonnet:2000}, 
no continuous or quasi continuous coupling of
atoms coherently into a guide was yet acheived until our realization of a guided
quasicontinuous atom laser \cite{Guerin:2006}. 

\begin{figure}[htb]
    \centering
    \includegraphics[width=10cm]{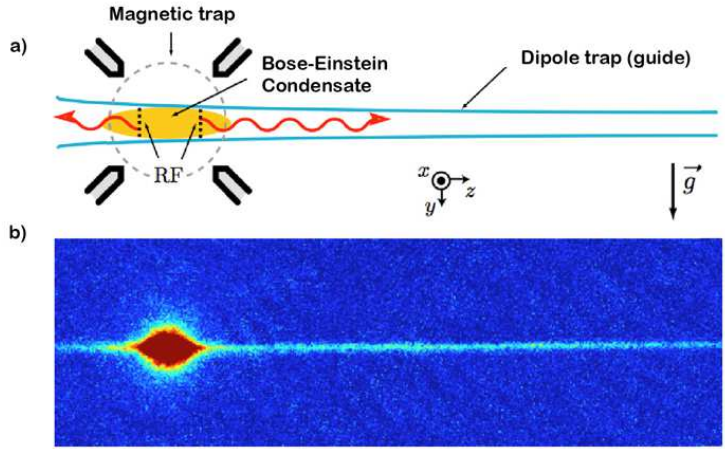}
    \caption{(a) Schematic view of the setup.
    The BEC is produced at the intersection of a magnetic trap and a
    horizontal elongated optical trap acting as an atomic waveguide
    for the outcoupled atom laser. An ``rf knife" provides outcoupling into the waveguide and an atom laser is
    emitted on both sides.
    (b) Absorption image (along $x$) of a guided atom laser after 100~ms
of outcoupling.} \label{fig_manip}
\end{figure}

In our realization of a guided
 atom laser \cite{Guerin:2006}, the coherent source,
\textit{i.e.} the trapped BEC, and the guide are {\em merged}
together in a hybrid combination of a magnetic Ioffe-Pritchard
trap and an horizontally elongated far off-resonance optical trap
constituting an atomic waveguide (see Fig. \ref{fig_manip}). The
BEC, in a state sensitive to both trapping potentials, is
submitted to a rf outcoupler yielding atoms in a state sensitive
only to the optical potential, resulting in an atom laser
propagating along the weak confining axis of the optical trap. In
addition to canceling the effect of gravity, this configuration
has several advantages. 
\begin{itemize}
\item Coupling into a guide from a BEC
rather than from a thermal sample allows us
to couple a significant flux into a small number of transverse
modes of the guide. Hence, in spite of the
lensing effect due to the interaction of the atom laser with the
trapped BEC \cite{lecoq:2001,Riou:2006}, adiabatic transverse mode
matching results into the excitation of only a small number of
transverse modes, and we discussed the possibility of achieving
single transverse mode operation of the atom laser in  \cite{Guerin:2006}.
\item Using a rf outcoupler rather than
releasing a BEC into a guide
results into quasicontinuous operation. 
\item  It is possible to
compensate the weak longitudinal trapping potential of the guide
by the antitrapping potential due to the second order Zeeman
effect acting onto the outcoupled atoms, resulting in an atom
laser with a quasiconstant large de~Broglie wavelength ($> 1$~$\mu$m
over 0.1~s of propagation in this work). 
\item Changing the
frequency of the outcoupler allows one to \emph{tune} the value of
the de~Broglie wavelength of the atom laser. One can also vary the atom-laser density from the
interacting regime to the noninteracting regime, offering the prospect to study linear as
well as nonlinear atom optics phenomena. 
\end{itemize}
We finally present our first investigations on the development of guided atom optical elements using light microstructures to reflect or transmit the atomic beam.

\section{Experimental set-up}

Our setup \cite{fauquembergue:2005,Guerin:2006} produces magnetically trapped
cold clouds of $\mathrm{{}^{87}Rb}$ in the
$\left|F,m_{F}\right\rangle=\left|1,-1\right\rangle$ state. During
the evaporative cooling, an optical guide produced by 120~mW of
Nd:YAG laser ($\lambda\!=\!1064$~nm) focussed on a waist of
30~$\mu$m is superimposed along the $z$ direction and
Bose-Einstein condensation is directly obtained in the
optomagnetic trap. In this hybrid trap, the optical guide ensures
a tight transverse confinement, with oscillation frequencies
$\omega_{x,y}/2\pi\!=\!\omega_{\perp}/2\pi\!=\!360$~Hz, large
compared to the frequencies characterizing the magnetic
confinement along the same axes ($\omega_x^m/2\pi\!=\!8$~Hz and
$\omega_y^m/2\pi\!=\!35$~Hz). In contrast, the longitudinal
confinement of the BEC along the $z$ axis is due to the shallow
magnetic trap with an oscillation frequency
$\omega_z^m/2\pi\!=\!\omega_{\mathrm{m}}/2\pi\!=\!35$~Hz. A BEC of
10$^5$ atoms has then a chemical potential
$\mu_\mathrm{BEC}/h\simeq 3.2$~kHz and Thomas-Fermi radii
$R_z\!=\!25~\mu$m and $R_\bot\!=\!2.4~\mu$m. The guided atom laser
is obtained by rf-induced magnetic transition \cite{Bloch:1999}
between the $\left|1,-1\right\rangle$ state and the
$\left|1,0\right\rangle$ state, which is submitted to the same
transverse confinement due to the optical guide, but is not
sensitive (at first order) to the magnetic trapping. We thus
obtain a quasicontinuous guided coherent matter wave freely
propagating along the optical guide [Fig. \ref{fig_manip}(b)].
This configuration, where the optical guide dominates the
transverse trapping of both the source BEC and the atom laser,
enables to collect the outcoupled atoms into the optical guide
with 100$\%$ efficiency.

\section{A large de Broglie wavelenght atom laser}

\begin{figure}[htb]
    \centering
    \includegraphics[width=10cm]{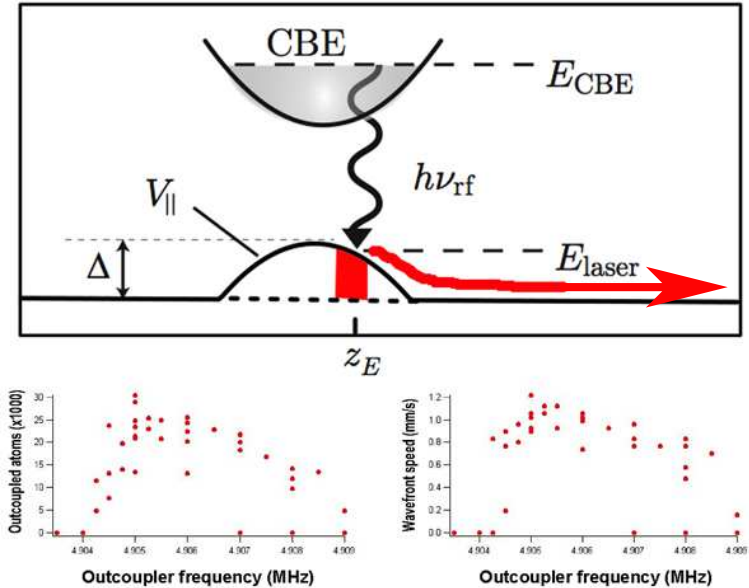}
    \caption{{\bf above} : Principle of RF outcoupling. Because atoms  
    can only be outcoupled (see \cite{lecoq:2001}) from where the resonance condition is fulfilled,
    fixing the RF frequency fixes the initial density and flux (and thus the amount of interaction energy) that is transferred to the atom laser.
    {\bf below} : Coupling curve (left) and corresponding velocity (right) with respect to the RF outcoupling frequency.} \label{fig_coupler}
\end{figure}

Taking advantage of having a horizontal guide, we can control the acceleration of the beam, which is not dominated
by gravity any more, as for usual atom laser experiment (see  \cite{Guerin:2006} for additional explanaitions). 
This is of great interest to obtain a coherent beam with
a large de~Broglie wavelength. The combination of the expulsive Zeeman quadratic effect and of the dipole
potential, whose center is shifted from $z_{0}$, results in a nearly linear potential whose slope is $m
\omega_{z}^{2} z_{0}$. Typical experimental parameters are $\omega_{z}=2\pi\times 2.5$~Hz and $z_{0}=1$~mm. This
gives an acceleration of $0.25$~m.s$^{-2}$, which corresponds to a reduction by a factor 40 compared to gravity.
Moreover, we can tune this acceleration by changing the power of the guiding laser, thus changing $\omega_{z}$, or
the position of the waist $z_{0}$ \cite{Guerin:2006}. Ultimately, we can cancel the residual acceleration so that
 the movement of the atoms is dominated by an initial velocity ``kick'' induced by mean-field interactions (see fig. \ref{fig_coupler}). 
Indeed, as can be seen on the diagram of fig. \ref{fig_coupler}, the outcoupled atoms experience first the
repulsive mean-field potential, which gives the beam an initial kinetic energy of the order of the chemical
potential $\mu(z_E)$ where $z_E$ is the position from which the atoms are extracted. We have studied this effect by measuring
the atom laser wavefront velocity (the method used and explained in \cite{Guerin:2006}) with respect to the RF outcoupler position
by changing the outcoupler frequency \cite{lecoq:2001,Riou:2006}.
 Changing this position changes the local density which is extracted and thus modulates the atom laser flux as shown in 
 fig.\ref{fig_coupler}. It also changes the initial velocity ``kick'' induced by the local mean field energy. 
 As shown in the preliminary studies on fig.\ref{fig_coupler},  
 we can change this velocity between $0.4$ mm/s and $1.2$ mm/s, corresponding to de Broglie wavelength between 4 and 12 $\mu$m, well
 above the de Broglie wavelength expected from ``classical'' energy considerations\footnote{The velocity kick resulting from mean field is expected to be around 5 mm/s.
 The lower velocity, that could be explained by energy spread over few transverse modes, is still under investigations.}.

\section{Towards a Fabry-Perot cavity for the atom laser}

Among the many possibilities opened up by the guided atom laser, the atom optical analog of a Fabry Perot cavity comes immediately to mind \cite{wilkens:1993}. 
In our experiment,  two potential barriers will be created by a laser beam at 405 nm in the TEM01 mode (with a node line in the center). 
The short optical wavelength is chosen to render the barriers sufficiently thin that the tunnelling amplitude cannot be prohibitively small. 
When the barriers are kept sufficiently stable in height and position, one expects the build up of a wave in the space between them. 

Our laser source is a 405 nm fibered diode laser delivering 30 mW of maximum power. The outpout of the fiber is collimated and a 2 mm diameter beam is sent 
to an anamorphic telescope, creating an anisotropic beam ($20\times 0.5$ mm$^2$) which is focused onto the atom laser with a numerical aperture of 0.15. 
This allows to create optical structure of width of the order of 1.3 $\mu$m FWHM along the propagation direction and 20 $\mu$m in the transverse direction. 
We calibrated our optical structure by sending clouds of atoms above and below BEC threshold onto a single optical barrier. 
This allows to measure the energy distribution within the cloud since in this ``classical'' regime, only the atom with kinetic energy higher than the barrier can be transmitted. 
With this 'spectroscopic' technique, we could record spectra of an atomic sample above and below BEC threshhold and demonstrate the control and stability 
of our barrier to better than 1 \% (100 Hz).

\begin{figure}[htb]
    \centering
    \includegraphics[width=11cm]{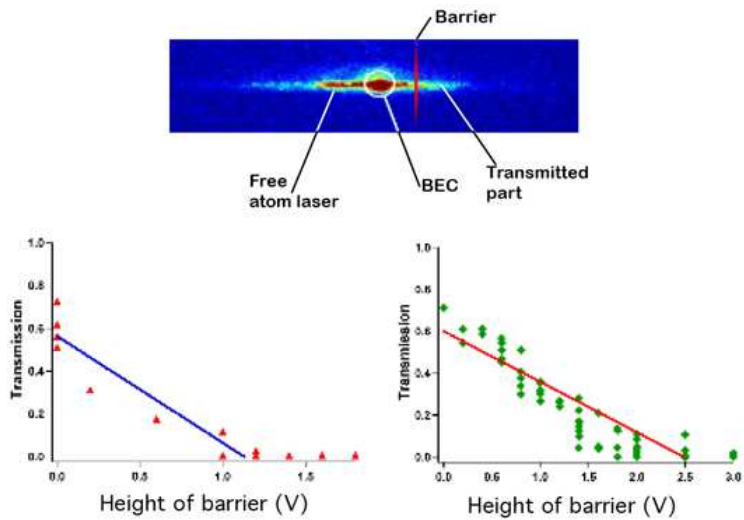}
    \caption{{\bf above} A dual beam atom laser used to study the transmission through a light barrier. The left side is used to calibrate the flux in a single shot 
    while the barrier is applied on the right. {\bf below} The transmission curves versus
    the laser intensity (the height in V is the amplitude control command, and correspond roughly to 2 kHz/V) for two different atom laser 
    energy (with estimated de Broglie wavelength of $\sim 1 \mu$m (left) and $\sim 1.5 \mu$m (right).
    }
    \label{Galpic}
\end{figure}

We now switch to our preliminary study of the transmission of the guided atom laser through the barrier. 
For that, first create a BEC in our hybrid trap as in \cite{Guerin:2006} and turn on the RF outcoupler for 100 ms.  By choosing the proper configuration, 
it is possible to create two atom lasers propagating in opposite directions from two symetric positions in the BEC. We set the optical barrier on one side of the BEC 
so that only one atom laser will be affected. This allows for a single shot calibration of the transmission factor through the barrier. 
The result for two different RF outcoupler frequencies, with estimated corresponding de Broglie wavelength of $\sim 1 \mu$m and $\sim 1.5 \mu$m are 
shown in figure \ref{Galpic}.
 It is interesting to see that a smooth transition occurs from total transmission (at 0 V) to total reflexion (at maximum barrier height). For an
 ideal atom laser, we expect nonetheless a step function since the energy spread should be negligible. This difference could be explained by unwanted broadening of the atom laser due to technical noise. We estimate this noise from the resonance curve of figure \ref{fig_coupler} to be below 1 kHz, to low to explain the transmission curve. Other explanations, such as more fundamental broadening due to residual excitation in the coupling process, are under investigation. A last explanation could come from the fact that the wavelength of the laser is of the order or bigger than the width of the barrier. In this case, tunneling is also expecting to broaden the transmission curve.

\section{conclusion}
We  present the realization of the first guided atom laser \cite{Guerin:2006}. Unlike other
atom lasers, in which the atoms accelerate under gravity after being outcoupled from a BEC in a trap, our
laser beam can be made to propagate horizontally and thus free of any acceleration. The atomic velocity
can be chosen by the outcoupling parameters and then the matter wave propagates with a constant de
Broglie wavelength (typically of order 1 $\mu$m or more) over several mm. One can also control the atomic density in
the beam and thus the amount of interaction between atoms. It becomes possible to explore various
regimes from non-interacting to strongly interacting beams. This type of laser beam is ideal for the study of transport phenomena, especially those related to
tunneling and quantum reflection in which low energies and weak interactions are favorable \cite{Pasquini:2006}. 
Among the many possibilities opened up by the guided atom laser, the atom optical analog of a
Fabry Perot cavity comes immediately to mind.
Nevertheless, by contrast to the typical optical case, the atoms interact. Thus any atom laser 
Fabry-P\'erot cavity will resemble an optical cavity containing a Kerr medium. The high confinement in this configuration means that the resonance energies 
are strongly shifted by the presence of other atoms, rendering the cavity bistable and producing a profound modification of the statistical properties of the atoms. 
Depending on the interaction strength, which is quantified by the maximum number of atoms $N_{\rm max}$
allowed in the cavity at the bistability threshold, different behaviors are expected for the quantum
properties of the transmitted atomic beam. As has been discussed \cite{Collett:1984,Reynaud:1989} and observed
in the optical domain \cite{Raizen:1987, Lambrecht:1996}, squeezing appears in the weak interacting regime
($N_{\rm max}\geq 1$), where the presence or absence of one atom does not significantly perturb the dynamics. Such
an observation will constitute in itself a pioneering experiment in the context of quantum atoms optics.

\end{document}